\documentclass[journal]{IEEEtran}
\usepackage{cite}
\usepackage{amsmath,amsfonts}
\usepackage{accents}

\usepackage{algorithmic}
\usepackage{array}
\usepackage{textcomp}
\usepackage{stfloats}
\usepackage{url}
\usepackage{verbatim}
\usepackage{graphicx}
\usepackage{lipsum}
\newtheorem{algorithm}{Algorithm}
\usepackage[utf8]{inputenc}
\usepackage{color}
\usepackage{amsmath}
\usepackage{amsthm}
\usepackage{amssymb}
\usepackage{graphicx}
\usepackage[T1]{fontenc}
\usepackage[active]{srcltx}
\usepackage{color}
\usepackage{float}
\floatstyle{ruled}
\newfloat{algorithm}{tbp}{loa}
\providecommand{\algorithmname}{Algorithm}
\floatname{algorithm}{\protect\algorithmname}

\ifCLASSOPTIONcompsoc
    \usepackage[caption=false, font=normalsize, labelfont=sf, textfont=sf]{subfig}
\else
    \usepackage[caption=false, font=footnotesize, labelfont=rm, textfont=rm]{subfig}
\fi

\usepackage{amsmath,graphicx}
\usepackage{amssymb}
\usepackage{amsbsy}\usepackage{epsfig}
\usepackage{cite}
\usepackage{balance}
\usepackage{xcolor}
\usepackage{float}
\floatstyle{ruled}
\newcommand{\algalign}[2]

\def\bu{\mbox{\boldmath $u$}}

\def\bx{\mathbf{x}}

\def\bs{\mbox{\boldmath $s$}}

\def\b1{\mbox{\boldmath $1$}}
\def\b0{\mbox{\boldmath $0$}}

\def\my{\mbox{$\mathbf{y}$}}

\def\mp{\mbox{$\mathbf{p}$}}

\def\mB{\mbox{$\mathbf{B}$}}

\def\mC{\mbox{$\mathbf{C}$}}
\def\mLambda{\mbox{$\mathbf{\Lambda}$}}

\def\mD{\mbox{$\mathbf{D}$}}

\def\mF{\mbox{$\mathbf{F}$}}

\def\mI{\mbox{$\mathbf{I}$}}

\def\mL{\mbox{$\mathbf{L}$}}

\def\mQ{\mbox{$\mathbf{Q}$}}

\def\mU{\mbox{$\mathbf{U}$}}
\def\mV{\mbox{$\mathbf{V}$}}

\def\mW{\mbox{$\mathbf{W}$}}

\def\mf{\mbox{$\mathbf{f}$}}
\newcommand{\ds}{\displaystyle}

\floatstyle{ruled}

\begin{document}
\title{Physics-Informed Topological Signal Processing for  Water Distribution Network Monitoring}
\author{Tiziana~Cattai, Stefania~Sardellitti, Stefania~Colonnese, Francesca~Cuomo, Sergio~Barbarossa

\thanks{T. Cattai, S. Colonnese, F. Cuomo and S. Barbarossa  are with the Department of Information Engineering, Electronics, and Telecommunications, Sapienza University of Rome, 00184 Rome, Italy (e-mail: name.surname@uniroma1.it); S. Sardellitti is with the Dept. of Engineering and Sciences, Universitas Mercatorum, Rome, Italy (e-mail: stefania.sardellitti@unimercatorum.it).}
\thanks{This work was supported by the European Union under the Italian National Recovery and Resilience Plan (NRRP) of NextGenerationEU, partnership on "Telecommunications of the Future" (PE00000001 - program “RESTART”).}
}
\maketitle

\begin{abstract}
Water management is one of the most critical aspects of our society, together with population increase and climate change. Water scarcity requires a better characterization and monitoring of Water Distribution Networks (WDNs).
This paper presents a novel framework for monitoring Water Distribution Networks (WDNs) by integrating physics-informed modeling of the nonlinear interactions between pressure and flow data with Topological Signal Processing (TSP) techniques. We represent pressure and flow data as signals defined over a second-order cell complex, enabling accurate estimation of water pressures and flows throughout the entire network from sparse sensor measurements.  By formalizing hydraulic conservation laws through the TSP framework, we provide a comprehensive representation of nodal pressures and edge flows that incorporate  higher-order interactions captured through the formalism of cell complexes. This provides a principled way to decompose the water flows in WDNs in three orthogonal signal components (irrotational, solenoidal and harmonic). The spectral representations of these components inherently reflect the conservation laws governing the water pressures and flows.
Sparse representation in the spectral domain enable  topology-based sampling and reconstruction of nodal pressures and water flows from sparse measurements. Our results demonstrate that employing cell complex-based signal representations enhances the accuracy of edge signal reconstruction, due to proper modeling of both conservative and non-conservative flows along the polygonal cells. 
\end{abstract}


\IEEEpeerreviewmaketitle

\section{Introduction}
\begin{figure*}[t]
    \centering{\includegraphics[scale=0.8]{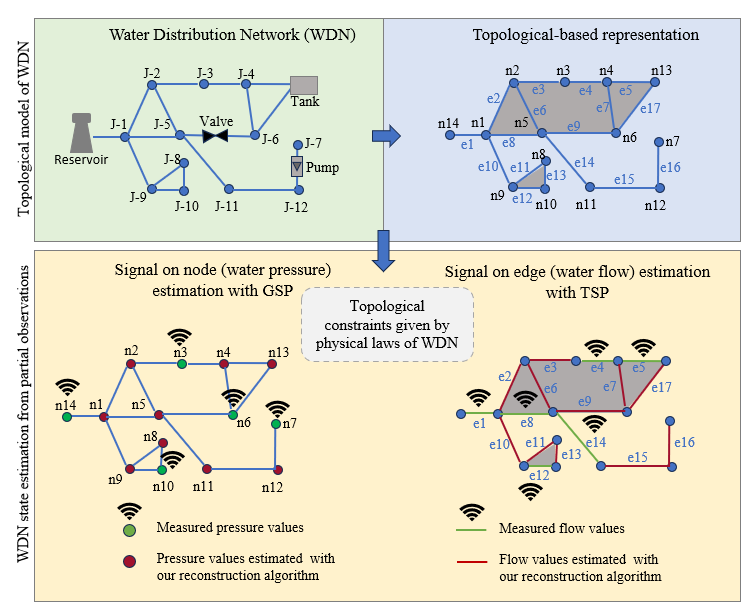}}
    \caption{Illustrative scheme of the proposed framework for WDN state estimation from partial observations.}
    \label{fig:schemamethod}
\end{figure*}

The increasing global challenge of water scarcity, driven by growing population demands and the effects of climate change, has intensified the need for efficient management of Water Distribution Networks (WDNs) \cite{liu2017water,pagano2024swi}. The precise estimation of pressure and flow within these networks in the presence of limited available sensor measurements is crucial for ensuring a reliable and sustainable water supply \cite{tzanakakis2020water, cattai2024graphsmart}. However, the intrinsic complexity of WDNs, together with the variability in data collection methods, poses significant challenges to traditional monitoring approaches, which often rely on manual inspections or rudimentary models. These conventional methods are not only labor demanding and also subject to inaccuracies \cite{liu2013state}.

In this context, water sensors can be strategically placed throughout the WDN to optimally monitor physical quantities of interest and transmit these data to a central system or device \cite{cattai2024graphsmart, ponti2021new,cheng2023optimal,wei2019optimal}. However, budget constraints and physical inaccessibility in some areas often limit the number of sensors and their possible deployment, making it impossible to place IoT sensors everywhere. As a consequence, a balance between cost-efficiency and thorough coverage of the WDN is needed. 

In this paper, we propose a novel approach that models pressure and flow data as signals defined over a second order cell complex and then leverages Topological Signal Processing (TSP) tools to model and analyze these data on high-order network structures \cite{sardellitti2024topological, barbarossa2020topological}.  Unlike graph-signal processing methods, which focus on signals defined over the vertices of a graph and are limited to capturing only pairwise interactions between nodes \cite{sandryhaila2013discrete}, TSP over cell complexes provides a more general framework for analyzing and processing signals defined over vertices {\it and} edges of a higher order structure, such as a cell complex, capable of exploiting higher order interactions \cite{sardellitti2024topological, barbarossa2020making}. Graphs are in fact a simple case of a cell complex, able to  encode only pairwise interactions.
The (higher order) topological perspective is particularly advantageous for accurately characterizing the interdependencies between flows and pressures  in a water distribution network. In this work, we  model  the topology of a WDN as a second order cell complex, i.e. as a topological space composed of nodes, edges and polygonal cells. This structure is well-suited for representing nodal pressures and flows as node and edges signals, respectively, defined over these topological domains. Interestingly, the use of cell complexes enables a richer representation of the interactions within the network, capturing both local and higher-order relationships. We  show how the Hodge spectral representations of these signals inherently encode the physical principles governing WDNs, specifically the mass conservation and flow conservation laws.
Recently, topological representation of WDNs networks through graphs  have been proposed in 
\cite{KERIMOV2025122980, money2024evolution,sabbaqi2024inferring}. In \cite{KERIMOV2025122980}, the authors formulate the problem of WDN state estimation as a diffusion process on the edges of a graph. 
Data-driven methods based on graph neural networks have been developed in recent years for WDNs state estimation \cite{ bentivoglio2024multi, kerimov2023assessing,xing2022graph}. In  \cite{gardharsson2022graph, leonzio2024water} the authors proposed ML-based and GNN-based methods to detect and classify types of leakage in WDNs.

Our goal in this paper is to develop topology-based learning strategies enabling the reconstruction of the vector states of WDNs,  as  nodal pressures and water flows, from a limited number of  sensors observations. 
We firstly introduce a topological representation of pressures and water flows that combines aspects of topological signal processing with the physical laws governing water flow in a WDN, incorporating nonlinear interactions between pressures and flows due to the hydraulic resistance present in each pipe.. This {\it physics-informed} representation allows us to analyze water flows and pressures in WDNs by intrinsically adhering to their topological and physical constraints. This model provides a new way to estimate the state of the overall WDN from a limited number of observations over a subset of nodes/edges.




To identify the optimal locations and the minimum number of necessary sensors, we employ state-of-the-art algorithms developed for node signals \cite{tsitsvero2016signals} and generalize them to flow (edge) signals. Unlike existing approaches in the literature that often rely on traditional or graph-based machine learning approaches, which often require large training datasets and a significant amount of computational resources, we adopt a model-based approach \cite{gardharsson2022graph}.
In addition, we integrate active elements within our framework, such as pumps, valves, and reservoirs, which are often overlooked in the available literature \cite{cattai2024graphsmart}. \\
Specifically, we first address the case where the signals are defined on the nodes of the graph, representing pressure values that can be measured by IoT devices positioned at the network junctions corresponding to the graph nodes. 
We also tackled the challenges arising from the nonlinear relationship between flow and pressure, as well as the presence of water demands, by formulating appropriate optimization problems. Then,  we design and test on realistic data, an optimization algorithm to estimate the pressures at unobserved points in the network by solving a non-convex optimization problem. We address the non-convexity of the formulated problem by iteratively solving a sequence of strongly convex problems converging to a local optimal solution.
Furthermore, we consider the case where the available signal represents the water flow measured at the edges of the graph. In such a case, we reconstruct the edge signals from the observed samples by solving a convex optimization problem that balances the data estimation error with the mass conservation principle. The formulated problem admits a closed-form optimal solution.
We then test our method on different realistic WDNs including water demands, efficiently reconstructing the flow across the entire network. Our results show that incorporating complex relationships through cell complex-based representations improves the accuracy and performance of flow reconstruction with respect to graph-based representations. An illustrative scheme of our framework is reported in Fig.\ref{fig:schemamethod}.\\
This paper is structured as follows: Section \ref{sec:tps} introduces the theoretical foundations of topological signal processing, focusing on graph signal processing and algebraic representation of cell complexes. Section  \ref{sec:topflow} presents the spectral representation of edge signals and their Hodge decomposition. In Section \ref{sec:topflow} the conservation laws governing the WDNs are reformulated in terms of the divergence-free and curl-free properties of edge signals.  In Sections \ref{sec:res} and \ref{sec:TSP_learn},
we formulate our optimization problems for reconstructing node pressures and flows as signals defined over graph nodes and edges, respectively. Furthermore, we  validate our approach with numerical experiments on realistic water distribution networks. Finally, we present our conclusions in Section \ref{sec:concl}.

\section{Background on Topological Signal Processing}
\label{sec:tps}

In  this section, we present an overview of the basic topological signal processing (TSP) concepts and tools \cite{sardellitti2024topological,barbarossa2020topological},   which will be pivotal for designing a novel topology-based method to learn water pressures and flows in  water distribution networks. We begin by introducing graph signal processing (GSP) \cite{shuman2013} and then we will extend the analysis to signals defined over higher-order topological spaces such as cell complexes. 

 \subsection{Graph Signal Processing Tools}
 \begin{figure}
    \centering{\includegraphics[scale=0.7]{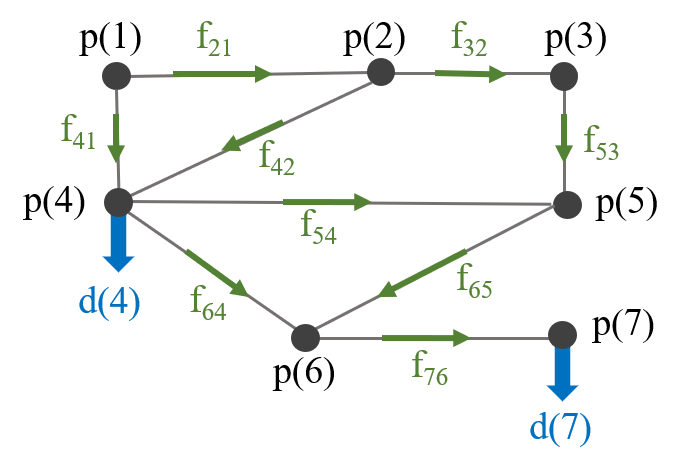}}
    \caption{An example of WDN graph with the observed pressures and demands as node signals.}
    \label{fig:schemagsp}
\end{figure}
As illustrated in Fig.  \ref{fig:schemamethod}  (upper left plot), a water distribution network can be efficiently modeled as a graph,  where junctions, reservoirs and storage tanks correspond to nodes, while  pipes, pumps and valves are represented as edges. This graph-based representation enables the  use of graph signal processing tools for the analysis of WDNs. GSP \cite{shuman2013} transposes classical signal processing operations, such as, for example, filtering, sampling, estimation, to signals defined over the vertices of a graph. \\ Let us  denote with $\mathcal{G}=(\mathcal{V},\mathcal{E})$ the undirected graph associated with a water network,   where  $\mathcal{V}=\{v_1,\ldots, v_N\}$  is the set of $N$ nodes  and $\mathcal{E}=\{e_{ij}\}_{i,j \in \mathcal{V}}$ is the set of edges with  $e_{ij}=1$, if there is a  link (pipe segment)  between node $i$ and node $j$, and $e_{ij}=0$ if the nodes are not connected. We denote by $E$ the number of edges in the graph, i.e.  $|\mathcal{E}|=E$. 
The connectivity relations can be represented by the incidence matrix $\mB_1$.n whose entries can be defined by first fixing an (arbitrary) orientation of the edges. The entries of $\mB_1$ are $B_1(i,j)=0$ if $e_{ij}=0$, $B_1(i,j)=1$ if node $i$ is the tail of the edge $e_{ij}$ and $B_1(i,j)=-1$ if node $i$ is instead the head of $e_{ij}$.
The Laplacian matrix is then defined as $\mL=\mB_1 \mB_1^T$ and is, by construction, a symmetric and positive semidefinite matrix. Its eigendecomposition can be written as $\mL= \mU \mLambda \mU^T$, where $\mU$ is the matrix whose columns are the eigenvectors $\{\bu_i\}_{i=1}^{N}$ of $\mL$, and $\mLambda$ is a diagonal matrix containing the associated eigenvalues.\\
A signal $\bx$ over a graph $\mathcal{G}$ is defined as a mapping from the vertex set to the set of real numbers, i.e., $\bx: \mathcal{V}\rightarrow \mathbb{R}$. For undirected graphs, the projection of $\bx$ onto the subspace spanned by the eigenvectors $\mU=\{\bu_i\}_{i=1}^{N}$ of the Laplacian matrix $\mL$, i.e., $\hat{\bx}=\mU^T \bx$  \cite{shuman2013}, is defined as the Graph Fourier Transform (GFT) $\hat{\bx}$ of the graph signal $\bx$. 
Given a set $\mathcal{K}$ of frequency indexes with $K=|\mathcal{K}|$, a    $\mathcal{K}$-bandlimited graph signal  is  a signal represented over  $K$ eigenvectors bases, i.e.
 $\bx=\mU_{\mathcal{K}} \hat{\bx}_{\mathcal{K}}$,
where $\mU_{\mathcal{K}}=\{\bu_i\}_{i\in \mathcal{K}}$ denotes the set of Laplacian eigenvectors over the frequency indexes $\mathcal{K}$, and $\hat{\bx}_{\mathcal{K}}$ are the associated graph Fourier coefficients. If a graph signal is bandlimited, i.e. it admits a sparse representation on the eigenvectors bases, using graph sampling theory, we can recover  the overall signal  from observations collected over a subset ${\cal S}$ of nodes with $|{\cal S}|<N$ \cite{tsitsvero2016signals}. Specifically, defining the sampled signal $\my=\mD_{\cal S}\, \bx$, with $\mD_{\cal S}$ a node-selection diagonal matrix whose $i$-th diagonal entry is  $1$ if $i \in {\cal S}$, and $0$ otherwise, we can recover, under the bandlimited assumption and the necessary and sufficient conditions established in \cite{tsitsvero2016signals}, the overall node signal $\bx$  through the following  formula \cite{tsitsvero2016signals}
\begin{equation}    
\label{eq:interpol}
\bx=  \big(\mI-\bar{\mD}_{\cal S} \mU_{\mathcal{K}} \mU_{\mathcal{K}}^T\big)^{-1} \my.
\end{equation}
with $\bar{\mD}_{\cal S}=\mI-{\mD}_{\cal S}$.\\
Therefore, as discussed in the ensuing sections, we can use this  sampling and recovering strategy  to estimate the nodal pressure in a WDN.
In such a network, both pressures and user demands are treated as node signals. As an example,  Fig.  \ref{fig:schemagsp}  illustrates the graph topology of a simple WDN consisting of $N=7$ nodes and $E=9$ edges.  Each node $i$ is associated with a pressure value, denoted by $p(i)$. Additionally, we consider a subset of nodes $\mathcal{V}_d \subset \mathcal{V}$, where the demand value $d(i)$ can be observed  $\forall i \in \mathcal{V}_d$, as represented  by blue arrows in the figure. Note that, the edge $e_{ij}$ connecting  nodes $i$ and $j$ is traversed by a flow $f_{ij}$, depicted by green arrows.

\subsection{Topological Signals Processing Tools}
GSP enables the representation and processing of signals defined over the vertices of graphs. However, graphs are only able to grasp pairwise relations between data associated with their vertices through the presence of edges.
To incorporate higher-order relations between data defined over the nodes and properly capture the relationships between flow signals, it is necessary to resort to higher order structures, such as simplicial or cell complexes.
Topological Signal Processing (TSP) has recently been introduced to analyze and process signals defined over simplicial or cell complexes \cite{barbarossa2020topological, sardellitti2024topological,barbarossa2020making}. 
Given a finite set $\mathcal{V}= \{ v_{i}\}_{i=0}^{N-1}$ of $N$ vertices, a {\it $k$-simplex} $\sigma^k_i$ is an unordered set of  $k+1$ points in $\mathcal{V}$.
A \textit{face} of the $k$-simplex
is a $(k-1)$-simplex. 
An \textit{abstract simplicial complex} ${\cal X}$ is a finite collection of simplices that is closed under inclusion of faces, i.e., if ${\cal \sigma}_i \in {\cal X}$, then all faces of $\sigma_i$ also belong to ${\cal X}$.
The order of a simplex is its cardinality minus one. 
Simplicial complexes handle relations of any order, but are constrained to respect
the inclusion property, stating that if a simplex belongs
to the space, all its subsets must belong to the space as well. However, in many applications, this constraint could be overly restrictive.
To overcome this limitation, we consider cell complexes, which do not need to satisfy the inclusion property. This enables the presence of elements having more intricate and sparse shapes, such as polygons with an arbitrary number of sides.\\ 
\begin{figure}    \centering{\includegraphics[scale=0.7]{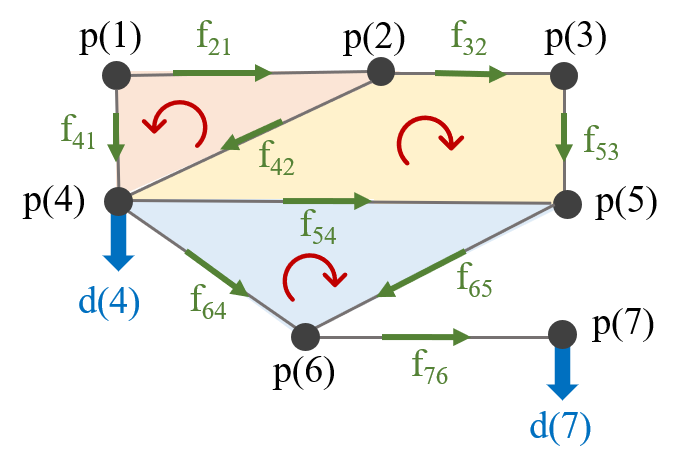}}
    \caption{Cell complex of order $2$ modelling  a WDN.}
    \label{fig:schemaCC}
\end{figure}
A cell complex (CC) consists of a collection of abstract elements (cells) characterized by a bounding relationship and a dimension function, which fulfill the properties of transitivity and monotonicity \cite{sardellitti2024topological}.
In this framework, an $n$-cell refers to a cell with dimension $n$:  vertices are considered $0$-cells, edges $1$-cells and polygons of any order are $2$-cells. 
In  Fig. \ref{fig:schemaCC}, we illustrate an example of a WDN modeled as a cell complex of order $2$ consisting of $N=6$ nodes, $E=9$ edges and $3$ cells of order $2$ (two triangles and one quadrilateral).\\
The boundary of an $n$-dimensional cell comprises all lower-dimensional cells that enclose it. A CC is a $K$-dimensional complex if all its cells have dimensions that are at most $K$.
According to these assumptions, simplicial complexes are specific types of cell complexes in which $2$-cells are triangles. The structure of a cell complex is represented by using its incidence matrices, i.e. a set of matrices detailing the relationship between $k$-cells and their corresponding $(k-1)$-cells. The orientation of a cell complex extends the notion used for simplicial complexes, encompassing the definition of orientation for cells of any order \cite{sardellitti2024topological}.\\
\textbf{Algebraic representation.}
Let us denote with $c_i^{k}$ the cell $i$ of order $k$. If  $c^{k-1}_i \prec_b c^{k}_j$ we say that $c^{k-1}_i$ is lower incident to  $c^{k}_j$. Two $k$-order cells are lower adjacent if they share a common face of order $k-1$ and upper adjacent if they are both faces of a cell of order $k+1$.
Given an orientation of the cell complex ${\cal C}$, the structure of a cell complex of order $K$ is fully captured by the set of its incidences matrices $\mB_k$ with $k=1,\ldots,K$, also named boundaries matrices, whose entries  establish which $k$-cells are incident to which $(k-1)$-cells. These boundary matrices are defined as:
  \begin{equation}      
   \label{inc_coeff}
  B_k(i,j)=\left\{\begin{array}{rll}
  0, & \text{if} \; c^{k-1}_i \not\prec_b c^{k}_j \\
  1,& \text{if} \; c^{k-1}_i \prec_b c^{k}_j \;  \text{and} \; c^{k-1}_i \sim c^{k}_j\\
  -1,& \text{if} \; c^{k-1}_i \prec_b c^{k}_j \;  \text{and} \; c^{k-1}_i \nsim c^{k}_j\\
  \end{array}\right. 
\end{equation}
where we use the notation $c^{k-1}_i \sim c^{k}_j$ to indicate that the orientation of $c^{k-1}_i$ is coherent with that of  $c^{k}_j$ and $c^{k-1}_i \nsim c^{k}_j$ to indicate that their orientations are opposite. To describe the structure of a $K$-cell complex we can consider the first order combinatorial Laplacian matrices 
\cite{Horak13} 
given by
\begin{equation}
\begin{split}
& \mL_0=\mB_1 \mB_1^T,\\
&\mL_k=\mB_k^T\mB_k+\mB_{k+1}\mB_{k+1}^T \; \; \mbox{for} \; k=1,\ldots,K-1\\
&\mL_K=\mB_K^T\mB_K
\end{split}
\label{eq:tsp}
\end{equation}
where $\mL_{k,d}=\mB_k^T \mB_k$ and $\mL_{k,u}=\mB_{k+1} \mB_{k+1}^T$ are  the lower and upper first-order Laplacians, expressing the lower and upper adjacencies of the $k$-order cells, respectively. An important property is that the boundary of a boundary is zero, i.e. it always holds $\mB_k \mB_{k+1}=\mathbf{0}$. Note that $\mL_0$ is the graph Laplacian matrix.
\section{Signal processing over cell complexes}
The fundamental tools for the analysis of signals defined over cell complexes are provided in \cite{sardellitti2024topological}. In this section, we briefly recall some properties that will be used later on. 
Let us consider w.l.o.g. a cell complex of order two $\mathcal{C}=(\mathcal{V},\mathcal{E},\mathcal{P})$
where $\mathcal{V}$, $\mathcal{E}$, $\mathcal{P}$ denote the set of $0$, $1$ and $2$-cells, i.e. vertices, edges and polygons, respectively. We denote their cardinality by $|\mathcal{V}|=N$, $|\mathcal{E}|=E$ and $|\mathcal{P}|=P$.
Then, the two incidence matrices describing the connectivity of the complex are $\mB_1 \in \mathbb{R}^{N\times E}$ and $\mB_2 \in \mathbb{R}^{E\times P}$, where $\mB_2$ can be written as
\begin{equation} \label{eq:B2_cell}
\mB_2=[\mB_{T},\mB_{Q},\ldots,\mB_{P_{max}} ]
\end{equation}
with  $\mB_{T}$, $\mB_{Q}$ and  $\mB_{P_{max}}$ describing the incidences between edges and, respectively, triangles, quadrilateral, up to polygons with  $P_{max}$ sides. The incidence matrix $\mB_2$ is then built identifying all cycles that are independent of each other, starting from triangles and then moving to higher order cells. In this way all independent polygons are identified.
\\
The eigenvectors of the Hodge Laplacian matrices have some interesting properties that are pivotal for the processing of topological signals. Specifically, let us consider w.l.o.g. the first-order Laplacian matrix  $\mL_1=\mB_1^T \mB_1+\mB_2 \mB_2^T$. Since, by construction, $\mB_1 \mB_2=\mathbf{0}$, it can be easily proven that the eigenvectors $\bu_i^{1}$ associated with the nonnull eigenvalues of $\mL_1$ are either the eigenvectors of $\mL_{1,d}:=\mB_1^T \mB_1$ or those of $\mL_{1,u}:=\mB_2 \mB_2^T$. Furthermore, the so called Hodge decomposition holds, stating that the vector space $\mathbb{R}^{E}$ of the edge vecotrs can be partitioned as:
\begin{equation} \label{eq:Hodge_dec}
\mathbb{R}^{E} \triangleq \text{img}(\mB_1^T) \oplus \text{ker}(\mL_1)\oplus \text{img}(\mB_2)
\end{equation}
where the vectors in $\text{ker}(\mL_1)$ are also in
$\text{ker}(\mB_1)$ and $\text{ker}(\mB_2^T)$.\\
Given a second-order cell complexes $\mathcal{C}=(\mathcal{V},\mathcal{E},\mathcal{P})$, the signals defined on the vertices, edges and polygons, represent maps $\bs^{0} : \mathcal{V} \rightarrow \mathbb{R}^{N}$, $\bs^{1} : \mathcal{E} \rightarrow \mathbb{R}^{E}$ and $\bs^{2} : \mathcal{P} \rightarrow \mathbb{R}^{P}$. From the Hodge decomposition in (\ref{eq:Hodge_dec}), it follows that an edge signal $\bs^{1}$ can be expressed as the sum of three orthogonal components \cite{Lim}:
\begin{equation} \label{eq: Hodge_dec}
\bs^{1}=\mB_1^T \bs^{0}+ \mB_2 \bs^2 + \bs^{1}_H.
\end{equation}
In analogy to vector fields, these three components of the flow satisfy conservative principles at the nodes or/and along the cycles of the complex. Specifically, the first component $\bs^{1}_{irr}:=\mB_1^T \bs^{0}$
represents the gradient, or irrotational, component, since  it has zero circulation along each $2$-order cell, i.e. 
\begin{equation} \label{eq:curl_free}    \mB_2^{T} \bs^{1}_{irr}=\mathbf{0}.
\end{equation}
Hence $\bs^{1}_{irr}$ is the component of the flow that is conservative  along the cycles of the network.
The second component in (\ref{eq: Hodge_dec}), $\bs_{sol}^{1}:= \mB_2 \bs^{2}$, is named the solenoidal signal, since its divergence, given by the sum of the flows in-going and out-going from each node, is equal to zero, i.e.
\begin{equation} \label{eq:div_free}
    \mB_1 \bs^{1}_{sol}=\mathbf{0}.
\end{equation}
Therefore, $\bs^{1}_{sol}$ is the component of the flow that is conservative at each node.\\
Finally, the last term in (\ref{eq: Hodge_dec}) represents the harmonic signal which lies in $\text{ker}(\mL_1)$  and  is conservative both at each node and along the cycles.\\
As discussed in \cite{sardellitti2024topological}, a useful basis to represent edge signals defined over cell complexes is given by the eigenvectors of $\mL_k$. Let us consider the eigendecomposition  of the Hodge Laplacian $\mL_k=\mU_k \boldsymbol{\Lambda}_k \mU_k^T$ where $\mU_k^T$  is the eigenvector matrix and $\boldsymbol{\Lambda}_k$ is a diagonal matrix with entries the associated eigenvalues. The cell Fourier Transform is defined as the projection of a $k$-order signal $\bs_k$ onto the eigenvectors of $\mL_k$, i.e., $\hat{\bs}_k=\mU_k^T \bs^{k}$ \cite{sardellitti2024topological}. A $k$-order signal can then be expressed in terms of its Fourier coefficients as $\bs^k=\mU_k \hat{\bs}_k$.


 \section{Topological representation of WDNs conservation laws}

 \label{sec:topflow}
 
In WDNs, physics-based models are used to describe and emulate the network behavior and estimate their states and parameters. These models, such as EPANET \cite{rossman2000epanet},   solve the mass and energy conservation equations to estimate the steady state of the system at any time in order to derive the flow rate on every pipe and the pressure on every node. In this section, we introduce the mathematical equations that we adopt to formulate our learning strategy.

\subsection{Conservation principles in WDNs}
Let us consider a WDN whose physical topology is modeled by a second-order cell 
complex  $\mathcal{X}=(\mathcal{V},\mathcal{E},\mathcal{P})$  composed by  $N$ nodes (junctions), $E$ edges and $P$ cells of order $2$.  The pressure measurements observed at each node are the nonnegative entries of the node vector $\mp_0\in \mathbb{R}^{N}_{+}$.
Denoting with  $f_{ij}$ the volumetric water flow rate measured  over each edge (pipe) $e_{ij}$, we introduce the flow vector 
$\mf \in \mathbb{R}^{E}$  with entries $f_{ij}$, $\forall \, e_{ij} \in \mathcal{E}$. 
The conservation laws governing  a WDN lead to two sets of equations \cite{todini2013} that describe the distribution of flows $\mf$ and pressures $\mp_0$ within a looped network of pipes.\\
The first set of equations is derived applying the \textit{conservation of mass law}  requiring that the total inflow equals the total outflow at each network node to which pipes are connected.  This implies that, assuming that the edge of the graph are oriented according to the water flow direction, we can impose this conservation law to each node $i$ as
 \begin{equation}
 \label{eq:con_law}
\ds \sum_{j \in \mathcal{N}_i^{\text{in}}}   f_{ij} =\ds \sum_{j \in \mathcal{N}_i^{\text{out}}} f_{ij}  \; \, \forall i \in \mathcal{V}
 \end{equation}
 where $\mathcal{N}_i^{\text{in}}$, $\mathcal{N}_i^{\text{out}}$ are the subset of the nodes for which there is a direct link in-going into the node $i$ and  out-going from node $i$, respectively. Note that the conservation mass flow at each node corresponds to the divergence-free condition given in (\ref{eq:div_free}).
 Hence, using the incidence matrix $\mB_1$, we can write equation (\ref{eq:con_law}) in matrix form  as 
 \begin{equation}
 \label{eq:con_law1}
\mB_1 \mf=\mathbf{0}.
 \end{equation}
Let us now consider the case where some nodes (junctions) may have demands, i.e. outgoing  flows from the node that turn it into a virtual sink. Denoting with  $\mathcal{V}_d$ the set of nodes where are observed  water demands, we define a nodal signal $\mathbf{d} \in \mathbb{R}^{N}$ with entries the nodal demand $d(i)$ for $i \in \mathcal{V}_d$  and zero otherwise. 
  Hence, we can rewrite (\ref{eq:con_law}) in matrix form as 
 \begin{equation}
 \label{eq:con_law2}
\mB_1 \mf=\mathbf{d}.
 \end{equation}
This equation imposes that the divergence of $\mf$  in a WDN without leakage is equal to the demands on the nodes. This implies that the irrotational  component $\mf_{irr}$ of the flow  is not zero.\\
The second physical principle governing the flow rate and pressure  in a WDN is the \textit{conservation of energy}  across the length of each pipe $k$ \cite{bhave1991analysis}. 
It states that the variation of the pressure at the endpoints of the edge depends  non-linearly on the edge flow as follows
\begin{equation} \label{eq:energ_cons}
p(v_i)-p(v_j)- \Phi(f_{ij})=0
\end{equation}
where $v_i$ and $v_j$ are the nodes at the endpoints of the pipe segment $e_{ij}=(v_i,v_j)$ and $\Phi(f_{ij})$ represents the head loss as a function of the flow rate $f_{ij}$. The  frictional head loss function is typically a non-linear function of $f_{ij}$ expressed as \cite{todini2013}
\begin{equation}
\Phi(f_{ij})= c_{ij} |f_{ij}|^{\alpha-1} f_{ij}  
\end{equation}
where $c_{ij}$ is a coefficient describing the hydraulic resistance of each pipe to the water propagation, while $\alpha$ is a coefficient which depends on the type of law that is used. Specifically, when using the  Darcy-Weisbach equation, we have $\alpha=2$ and,  for cylindrical pipes, we have $c_{ij}=\ds \frac{\mu\, l_{ij} \rho}{8 d^5 \pi^2}$ where $\mu$ is the friction coefficient, $l_{ij}$ the length of the pipe $(v_i,v_j)$ connecting junction $i$ with junction $j$, $\rho$ the fluid density and $d$ the diameter of the pipe (assumed here equal for all the pipe) \cite{hajgato2021}.
However, in many practical situations the Darcy-Weisbach equation is intractable as the friction coefficient is unknown. 
Hence, empirical formulas,  such as the Hazen-Williams formula (H-W), are commonly used. In this formula we have $\alpha=1.852$ and $c_{ij}=\frac{4.727 \, l_{ij}}{d^{4.871} r^{1.852}}$
where $r$ is the Hazen-Williams roughness coefficient of the pipe, determined empirically.\\
Note that the  variation $\Delta \mp=p(v_i)-p(v_j)$ of the nodal pressure in (\ref{eq:energ_cons}) can be equivalently derived as the irrotational component of the Hodge decomposition of an edge signal \cite{grady2010discrete}, i.e. by using the  gradient operator  $\mB_1^T$ applied to the nodal signal as
\begin{equation}
\label{eq:Delta_B1}
    \Delta \mp=\mB_1^T \mp.
\end{equation}
Then,  (\ref{eq:energ_cons}),  can be  easily written as
 \begin{equation}
 \label{eq:p0_Phi}
    \mB_1^T \mp=\mathbf{\Phi}(\mf)=\mC \, \mathbf{h}(\mf)
 \end{equation}
 where $\mC$ is a $E\times E$ diagonal matrix with entries $c_{ij}$ and $\mathbf{h}(\mf)$ is a vector with entries $|f_{ij}|^{\alpha-1} f_{ij}$.
 Furthermore,  from (\ref{eq:p0_Phi}), using the curl-free property of the  irrotational flow $\mB_1^T \mp$, we easily get that its  circulation along each cell of the complex is zero, so that it results
 \begin{equation}
 \label{eq:circ_0}
    \mB_2^T \mC \, \mathbf{h}(\mf) =\mathbf{0}.
 \end{equation}
 This last system of equations is the Kirchoff's second law according to which the total head loss around a closed loop of pipes must be zero. 
Then,  from (\ref{eq:con_law2}), (\ref{eq:p0_Phi})  and  (\ref{eq:circ_0}), we get  the system of equations to be solved, i.e. 
 \begin{equation}
 \label{eq:syst_1}
 \left\{
 \begin{split}
 & \text{(a)} \,\mB_1 \mf=\mathbf{d}  \medskip\\
    & \text{(b)}  \,\mB_1^T \mp=\mC \mathbf{h}(\mf) \medskip\\
    &  \text{(c)} \, \mB_2^T \mC \, \mathbf{h}(\mf) =\mathbf{0}.
    \end{split} \right.
 \end{equation}
Note that  if  the edge flow $\mathbf{f}$ and the nodal pressure $\mp$ are treated as unknowns, the system in (\ref{eq:syst_1}) is a non-linear system. This system  can be solved using different linearization methods, as discussed in \cite{todini2013}.

\begin{figure}[t]
\centering{\includegraphics[scale=0.7]{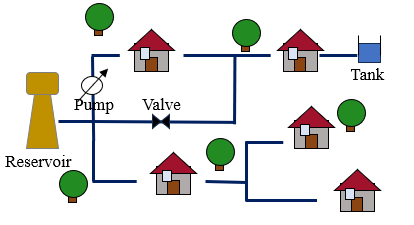}}
    \caption{Representation of a simplified WDN with active and passive elements. Active elements are reservoirs, pumps, valves and tanks while passive elements are junctions and pipes.}
    \label{fig:schemactive}
\end{figure}
\subsection{Active and passive elements in WDNs}
Thus far, we have described the behavior of the WDN under the simplifying assumption that all nodes and edges are homogeneous in  their functionalities  \cite{alperovits1977design,mala2017lost, cavalieri2024digital}. However, this is not the case in realistic WDNs, where specific active elements introduce additional complexities and must be appropriately accounted for in the analysis . These active elements include reservoirs, pumps, valves, and storage tanks, which differ significantly from passive elements such as pipes and junctions. A graphical representation of these elements in a simplified network is provided in Fig. \ref{fig:schemactive}.

Reservoirs, for example, are considered ideal sources of water, capable of supplying an infinite amount of flow. Pumps, on the other hand, facilitate the flow of water while consuming energy. The valves can be open, closed, or partially open to allow a specific amount of flow. Finally, tanks store water, contributing to the dynamical behavior of the network. For these components, the relationship between flow and pressure deviates from the standard equations derived for passive elements. In these cases, the flow is not solely determined by pressure differences, as assumed for regular pipes.

To correctly apply the conservation equations described earlier, it is necessary to exclude active elements from the graph representation or to include them in appropriate manner. For example, in the case of reservoirs, the EPANET simulator treats their contribution as a negative demand, meaning they inject water into the system rather than consuming it. Similarly, the behavior of pumps, valves, and tanks must be modeled separately to ensure accurate characterization of the network.\\
By appropriately taking into consideration active and passive elements and adjusting the mathematical model to account for these complexities, we ensure that our framework remains compatible with realistic scenarios, enabling accurate reconstruction of flow and pressure values within WDNs.

\section{GSP-based learning of nodal pressures}
\label{sec:res}
Water distribution networks  are critical infrastructure systems that are difficult to manage and monitor due to their size and complexity. Sensors revealing pressures or flows are sparsely deployed in the network, so critical challenges in WDNs are the recovery of the water quantities across all  nodes or edges. Our goal in this section is to develop a strategy to learn the optimal pressure values using a limited set of measurements.  
Our method will be tested on different realistic datasets in order to assess its effectiveness in  monitoring WDNs.\\
The proposed approach aims 
 to minimize the data-fitting error while ensuring compliance with the fundamental principles of flow conservation within the network.\\
 Let us denote with $\mathcal{S}$ the set  of nodes  where pressure sensors are placed. Then, we introduce the vector  $\bar{\mathbf{p}}_{\mathcal{S}}$  whose entries $p_{\mathcal{S}}(i)$ are the pressure measurements for $i \in \mathcal{S}$  and zero otherwise. Then,
we consider the system of equations (b) in  (\ref{eq:syst_1}). This system
 can be equivalently written as 
\begin{equation}
\label{eq:edg_flow}
    \mC^{-1} \mB_1^T \mp=\mathbf{h}(\mf)
\end{equation}
where the entries of the vector $\mathbf{h}(\mf)$  can be expressed for each edge $k$ as $h(k)=|f_k|^{\alpha-1} f_k=|f_k|^{\alpha} \text{sign}(f_k)$, with $k=1,\ldots,E$.  Let us now introduce the notations 
$|\mathbf{a}|^{ \circ 1/\alpha}$ and $\mathbf{a} \circ \mathbf{b}$  to denote, respectively, the Hadamard root and the Hadamard product, i.e.  a vector with entries  $|a|_k^{ 1/\alpha}$ and  $a_k b_k$.
Then,  from (\ref{eq:edg_flow}), we easily get
\begin{equation}
\label{eq:edg_flow1}
    |\mathbf{f}|= |\mC^{-1} \mB_1^T \mp|^{\circ 1/\alpha}.
\end{equation}
Replacing (\ref{eq:edg_flow1})  in the system (a) in (\ref{eq:syst_1}), we easily get 
\begin{equation}
\label{eq:edg_flow2}
    \mB_1 \mC^{-1/\alpha} (|\mB_1^T \mathbf{p}_0|^{\circ 1/\alpha} \circ \text{sign}(\mB_1^T \mathbf{p}))-\mathbf{d}=\mathbf{0}
\end{equation}
where $\mC^{-1/\alpha}$ denotes a $E\times E$ diagonal matrix with positive entries $c_{ij}^{-1/\alpha}$. \\
Assuming that the nodal pressure $\mathbf{p}$ is a $K$-bandlimited graph signal, we can express the pressure values  as $\mathbf{p}=\mU_{\mathcal{K}}\hat{\mathbf{p}}_{\mathcal{K}}$, where $\mU_{\mathcal{K}}$ contains as columns the eigenvectors of the graph Laplacian $\mL_0$ spanning the pressure graph signal. 
To find this sparse signal representation, we apply the method proposed in \cite{barbarossa2020topological} where the set of eigenvectors indexes $\mathcal{K}$ is derived by exploring the trade-off between the data fitting error and the sparsity of the representation.

Assume that  by using a water network simulator, such as EPANET, we can generate, according to the chosen WDN model,  the node pressure vectors  $\bx^0$. Given these vectors,   we   find a  sparse representation for the pressures vectors constrained to respect a maximum data fitting error  $\epsilon$,  as solution of the following basis pursuit problem
\cite{Donoho98}:
\begin{equation}
    \label{eq:bas_pur}
\begin{array}{lll} \underset{{\hat{\mathbf{p}}} \in \mathbb{R}^N}{\text{min}} & \parallel
\hat{\mathbf{p}}\parallel_1   \qquad \qquad \qquad (\mathcal{P}_B)\\
 \; \; \text{s.t.} & \parallel
 {\bx}^0 -\mV \hat{\mathbf{p}}\parallel_F \leq \epsilon
 \end{array}
\end{equation}
where $\mV$ is a dictionary matrix, whose columns are the eigenvectors $\mU$ of the Laplacian graph matrix $\mL_0$.
The solution of problem $\mathcal{P}_B$ returns the set of eigenvectors indexes $\mathcal{K}$, i.e. the eigenvectors matrix $\mU_{\mathcal{K}}$ spanning the pressures vectors $\mp$. To  derive the sampling set $\mathcal{S}$ where the sensors are placed to gather water flow measurements, we leverage the Max-Det method introduced for node signals in \cite{tsitsvero2016signals}.\\
Therefore, we can 
recover the overall pressures from the observation of a reduced number of  sampled measurements, by solving  the following optimization problem:
\begin{equation}
\label{eq:optim_prob}
\begin{array}{lll}
\underset{\mathbf{p} \in \mathbb{R}^{N}}{\min}  & \parallel \!(\mI-\bar{\mD}_{\mathcal{S}} \mU_{\mathcal{K}} \mU_{\mathcal{K}}^T)\mathbf{p}-\bar{\mathbf{p}}_{\mathcal{S}}  \parallel  + \\
&+\lambda \parallel \mB_1  \mC^{-1/\alpha} (|\mB_1^T \mathbf{p}|^{\circ 1/\alpha} \circ \text{sign}(\mB_1^T \mathbf{p}))-\mathbf{d} \parallel\medskip\\
\text{s.t.}  &  \mathbf{p} \geq \mathbf{0}  \hspace{5cm} (\mathcal{P})
\end{array}
\end{equation}
 where: i) the first term in the objective function quantifies the data-fitting error that arises when a graph signal is reconstructed from a limited subset of samples; ii) the second term ensures adherence to  the flow conservation principle within the WDN, controlled by  the non-negative penalty parameter  $\lambda \geq 0$. Finally, the constraint forces  the pressures to be non-negative. 
Unfortunately, problem $\mathcal{P}$ is a non convex problem due to the non-convexity of  the function
\begin{equation}
 g(\mathbf{p})=\parallel \mB_1  \mC^{-1/\alpha} (|\mB_1^T \mathbf{p}|^{\circ 1/\alpha} \circ \text{sign}(\mB_1^T \mathbf{p}))-\mathbf{d} \parallel.
\end{equation}
To solve the problem in \eqref{eq:optim_prob} efficiently, we apply a successive convex  approximation technique where  $\mathcal{P}$ is replaced by a sequence of strongly convex problems \cite{Scutari_17}. Therefore, we approximate the second non-convex term $g(\mathbf{p})$  in the objective function of $\mathcal{P}$ with a strongly convex approximation $\tilde{g}(\mathbf{p};\mathbf{p}^{\nu})$ around the iterate feasible point $\mathbf{p}^{\nu}$. The approximate strongly convex function is given by 
\begin{equation}
\label{eq:lin_g}
    \tilde{g}(\mathbf{p};\mathbf{p}^{\nu})=g(\mathbf{p}^{\nu})+\nabla^T_{\mp} g(\mathbf{p}^{\nu}) (\mp-\mp^{\nu})+\tau \parallel \mp-\mp^{\nu}\parallel^2
\end{equation}
where the last quadratic regularization term with $\tau>0$ is added to make the function strongly convex.
Defining  the matrix $\mQ:=\mC^{-1/\alpha} \mB_1^T \mB_1 \mC^{-1/\alpha}$, it holds
\begin{equation}
\label{eq:grad_g}
\begin{split}
    \nabla^T_{\mp} g(\mathbf{p})&=\left[(\mB_1 \circ \mF) \mQ (|\mB_1^T \mathbf{p}_0|^{\circ \frac{1}{\alpha}} \circ \text{sign}(\mB_1^T \mathbf{p}))\right.\\ 
   & \left.-\mB_1 \mD_G |\mB_1^T \mathbf{p}|^{\circ \frac{1}{\alpha}-1}\right] \frac{1}{\alpha  g(\mathbf{p})}
    \end{split}
\end{equation}
with $\mF$ a matrix composed by $E$ rows, each equal to $|\mathbf{p}^T\mB_1|^{\circ \frac{1}{\alpha}-1}$, and $\mD_G=\text{diag}(\mathbf{d}^T \mB_1 \mC^{-1/\alpha})$. We are now ready to introduce  the proposed convex approximation of the nonconvex objective function around the feasible point $\mathbf{p}^{\nu}$ as
\begin{equation}
\label{eq:grad_J}
\begin{split}
    \mathcal{J}\big(\mathbf{p}; \mathbf{p}^{\nu}\big)&:=\parallel \!(\mI-\bar{\mD}_{\mathcal{S}} \mU_{\mathcal{K}} \mU_{\mathcal{K}}^T)\mathbf{p}-\bar{\mathbf{p}}_{\mathcal{S}}  \parallel +\\
&+\lambda \cdot \nabla^T_{\mp} g(\mathbf{p}^{\nu}) (\mp-\mp^{\nu})+\tau \parallel \mp-\mp^{\nu}\parallel^2
\end{split}
\end{equation}
Therefore, we iteratively solve the following convex approximation of problem $\mathcal{P}$:
\begin{equation}
\label{minJ}
\begin{split}
    \hat{\mp}(\mp^{\nu}) :=\underset{\mathbf{p} \in \mathbb{R}^{N}}{\text{argmin}}  & \; \; \mathcal{J}\big(\mathbf{p}; \mathbf{p}^{\nu}\big) \hspace{1.7cm} (\mathcal{P}^{\nu})\\
\text{s.t.}  &  \; \;   \;\mathbf{p} \geq \mathbf{0}  
\end{split}
\end{equation}
where $\hat{\mp}(\mp^{\nu})$ is the unique solution of the strongly convex optimization problem $\mathcal{P}^{\nu}$.

   \begin{algorithm}[H]
\textbf{Initial  data:} $\mathbf{p}^0\in \mathbb{R}^{N}_{+}$; $\{\gamma^{\nu}\}_\nu \in (0,1]$; 

(\texttt{S.1}): If $\mathbf{p}^{\nu}$ satisfies a suitable termination criterion, \texttt{STOP}

(\texttt{S.2}): Compute    $\hat{\mathbf{p}}(\mathbf{p}^{\nu})$ as the solution of \eqref{minJ};

(\texttt{S.3}):  Set $\mathbf{p}^{\nu+1}=\mathbf{p}^{\nu}+\gamma^{\nu}\left(\hat{\mathbf{p}}(\mathbf{p}^{\nu})-\mathbf{p}^{\nu}\right)$;

(\texttt{S.4}):   $\nu \leftarrow \nu+1$  and go  to (\texttt{S.1}).

\caption{\textbf{:}  Inner SCA  Algorithm for  $\mathcal{P}$  \label{alg:Alg_centr}}

\end{algorithm}
The proposed method consists in solving the sequence of problem $\mathcal{P}^{\nu}$, starting from a feasible point $\mathbf{p}^0$. The final solution is a local optimal solution of $\mathcal{P}$ \cite{Scutari_17}.The formal description of the method is given in Algorithm $1$. Note that the optimal solution
 $\widehat{\mathbf{p}}$ of $\mathcal{P}^{\nu}$ computed in
Step 2 of the algorithm is used in Step 3 to set the  next iterate  $\mathbf{p}^{\nu+1}$
 by including a step-size in the updating rule. Many choices are possible for the step-size  $\gamma^{\nu}$; a practical rule is \cite{scutari_facchinei_et_al_tsp13}: \begin{equation}\gamma^{\nu+1}=\gamma^{\nu}(1-\bar{\alpha} \gamma^{\nu}), \quad  \gamma^{0}\in (0,1],
\label{eq:diminishing_step_size1} \end{equation} with $\bar{\alpha} \in \left(0,1/{\gamma^{0}}\right)$.

\begin{figure}
\centering{\includegraphics[scale=0.75]{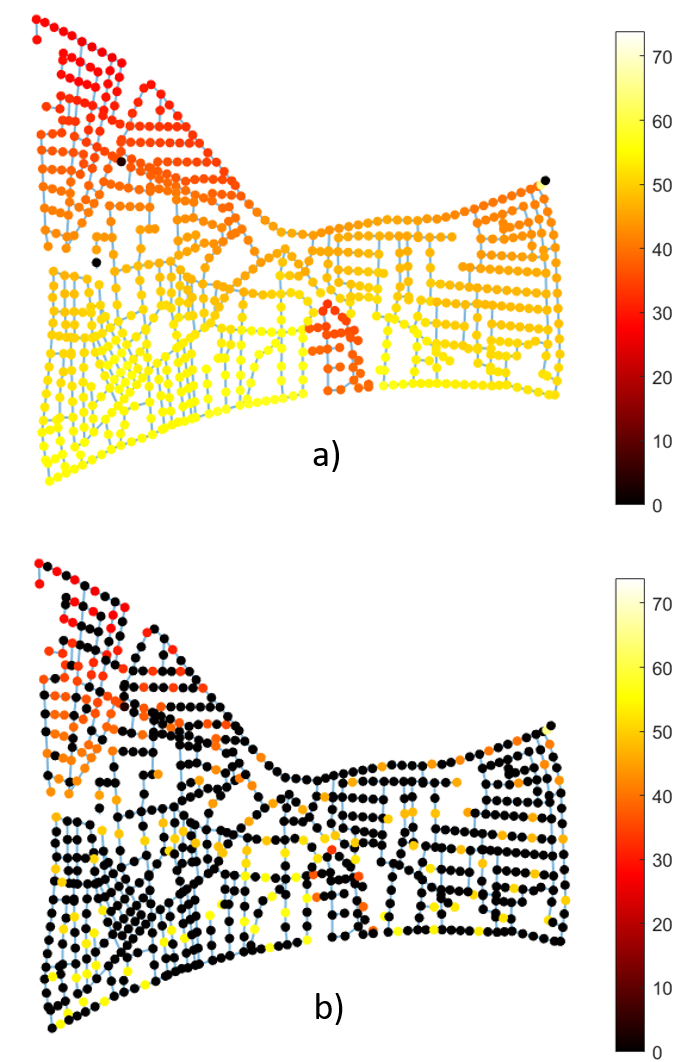}}
    \caption{In the upper plot a) we represent the node pressure values for the WDN in Cyprus. In the bottom plot b) we represent the pressure values, colored only for the sensors corresponding to the sampled set.}
    \label{fig:p}
\end{figure}

\noindent \textbf{Simulation results.} We test our method on a realistic water distribution network  \cite{Vrachimis2022}, whose topology comes from a modified version of the coastal city WDN in Cyprus, altered for security. This network contains $N=782$ junctions,  two reservoirs, one tank, $E=905$ pipes, one pump and two valves.  The model of the town was imported in EPANET software in order to simulate the water pressures and the hydraulic characteristics of the network \cite{rossman2000epanet}. 
We  model this WDN with a graph $\mathcal{G}$ with $N=785$ nodes and $E=909$ edges comprised of active and inactive elements.
An example of the pressure values encoded by the node colors, as obtained using EPANET, is given in Fig. \ref{fig:p}(a).
To select the subset of nodes where  sensors are placed to sample the nodal pressures, we used the Max-Det greedy sampling strategy  proposed in  \cite{tsitsvero2016signals}.
In Fig. \ref{fig:p}(b), we illustrate an example of sensor placement. The number of sampled nodes, is equal to  $187$.  Specifically, we represent in black the nodes that are not measured while in the color map we report the pressure values collected by the selected sensors.

\begin{figure}[t]
    \centering
    {\includegraphics[width=1\columnwidth]{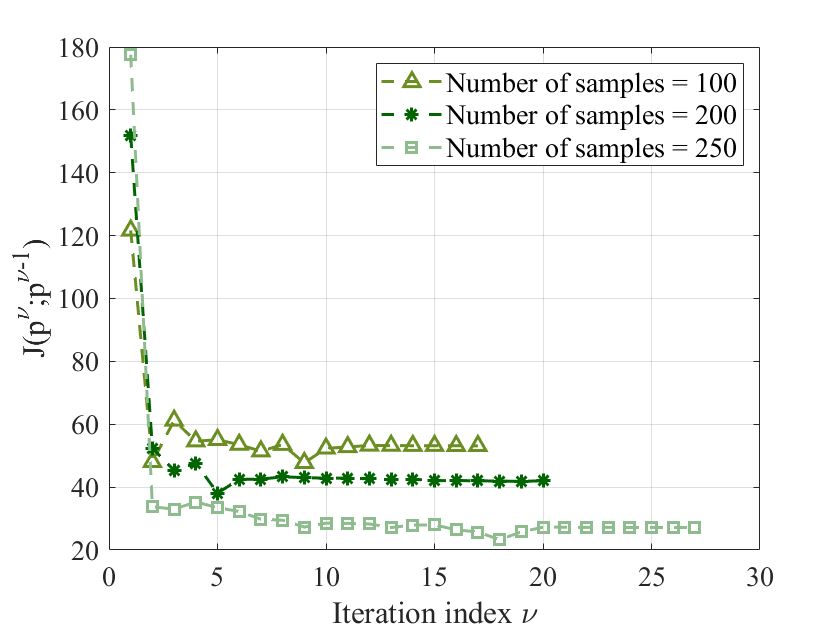}}
    \caption{The optimal cost function $J(\mathbf{p}^{\nu};\mathbf{p}^{\nu-1})$ versus the iteration index $\nu$ for different number of samples.}
    \label{fig:converg}
\end{figure}
\begin{figure}
    \centering{\includegraphics[scale=0.42]{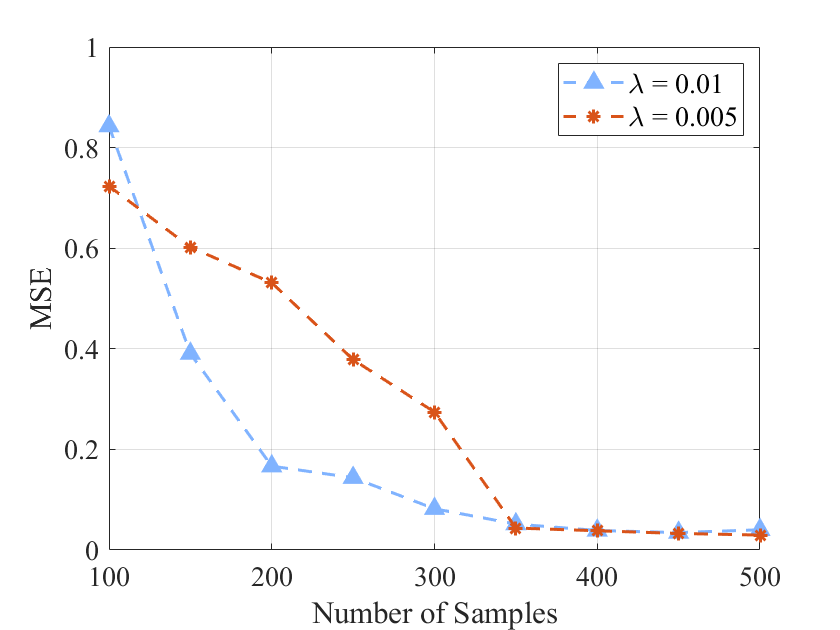}}
    \caption{Normalized MSE between reconstructed and ground truth pressure signals versus the number of samples and for different $\lambda$ values.}
    \label{fig:mse}
\end{figure}
Let us now  evaluate the convergence of the proposed SCA-based approximation method, which iteratively refines the solution by solving a sequence of strongly convex approximations of the original problem. The iterative procedure begins with an initial feasible point $ \mathbf{p}^{0}$, consisting of the sampled pressure values at the measured locations and random values for the unmeasured nodes. At each iteration, the estimation of the pressure value is derived according to Algorithm 1 until convergence is reached according to a prescribed accuracy. 
To evaluate the convergence of the proposed approach, we consider the evolution of the cost function \(\mathcal{J}(\mathbf{p};\mathbf{p}^{\nu-1})\) in (\ref{eq:grad_J}) versus the iteration indexes for different number of samples. In Fig. \ref{fig:converg} we can observe as  the cost function achieves a fast convergence in a few number of iterations. Furthermore, as expected, the minimum optimal value of the cost function decreases as the number of samples grows. 
To evaluate the goodness of the reconstruction strategy in \eqref{eq:optim_prob}, we consider the  Mean Square Error (MSE) between the ground truth and the reconstructed pressures normalized with respect to the norm of the signal. Then, we evaluate the MSE by varying the number of measured samples for different values of the regularization parameter $\lambda$.
For the experiments, the regularization parameter \( \tau \) in \eqref{eq:lin_g} was set to \( \tau = 0.2 \). The normalized Mean Squared Error (MSE) was computed at convergence, comparing the reconstructed pressure values against the ground truth, which was generated using EPANET simulations of a realistic water distribution network. The results, presented in Fig.~\ref{fig:mse}, show the normalized MSE as a function of the number of sampled nodes for different values of the regularization parameter \( \lambda \).  The results in Fig.\ref{fig:mse} illustrate the trade-off between reconstruction accuracy and the number of deployed sensors, as well as the impact of the regularization parameter on the reconstruction performance. Interestingly, as $\lambda$  increases, enforcing greater compliance of the estimated pressures  with the conservation law,  the recovery of the pressure values becomes more accurate.

\section{TSP-based learning of the  water flows}
\label{sec:TSP_learn}

\begin{figure}[ht]
    \centering
    {\includegraphics[width=1\columnwidth]{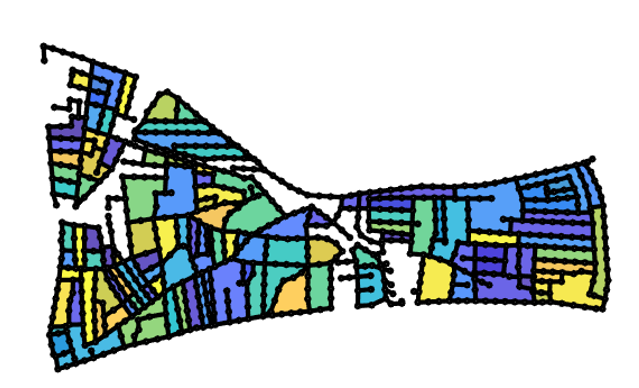}}
    \caption{Representation of the cell complex associated with the L-town  water distribution network. Each polygonal cell is highlighted in color.}
    \label{fig:cell}
\end{figure}

In this section, we consider the reconstruction of the water flows  from the observation of a reduced number of samples. Indeed in water distribution networks, sensors are often deployed to measure water flows, which are edge signals defined on the network.
However, the deployment of flow sensors across all edges of a water distribution network is often infeasible due to high installation and energy costs, as well as physical constraints that prevent sensors from being installed on every pipe. This limitation highlights the need for a robust reconstruction algorithm capable of estimating flow values across the entire network from a limited subset of sensor measurements. This approach is essential for effective monitoring of WDN, especially given the challenges posed by aging infrastructure and historically insufficient monitoring efforts.\\
Unlike node signals, such as pressures, which can be properly analyzed using a graph structure, when dealing with edge signals it is more appropriate to use a second-order cell complex, to properly capture, for example,  circulations of the water flows along the cycles of the network. For this reason, we model the physical topology of a WDN as a second-order cell complex. An example is shown in Fig. \ref{fig:cell}, where we sketch the topology of  the L-town WDN represented by a cell complex with polygonal cells having a maximum number of sides equal to $P_{max}=30$.  
Leveraging the theoretical developments in Section III, let us consider the first-order combinatorial Laplacian $\mathbf{L}_1$ given by :

\begin{equation}
    \begin{array}{lll}
        \mathbf{L}_{1} =  \mathbf{B}^{T}_{1} \cdot \mathbf{B}_1  + \mathbf{B}_2 \cdot \mathbf{B}^{T}_{2} = \mathbf{L}_{1,d} +  \mathbf{L}_{1,u}.
    \label{eq:L_1}
    \end{array}
\end{equation}
The lower Laplacian matrix  $\mathbf{L}_{1,d} =  \mathbf{B}^{T}_{1} \cdot \mathbf{B}_1 $  accounts for the node-to-edge incidence, 
while the upper Laplacian, $\mathbf{L}_{1,u} =  \mathbf{B}_{2} \cdot \mathbf{B}^{T}_2 $,  captures the incidence between edges and polygonal cells. 

To build the edge-polygons incidence matrix $\mB_2$ we leverage the method proposed  in \cite{sardellitti2024topological}, which 
systematically constructs all cycles in the graph, starting with a minimum cycle length of three (triangles) up to a predefined maximum length $P_{max}$. A new cyle is added as a column of $\mB_2$ only if it is linearly independent of the previous columns. These cycles are then oriented so that 
the resulting $\mathbf{B}_2$ matrix  encodes the relationships between edges and cells, with columns associated with filled polygonal cells, and  entries reflecting the orientation consistency between edge direction and cells. To ensure linear independence and numerical stability, the algorithm performs a column reduction step in order to ensure that matrix captures the minimal cycle bases needed for constructing $\mathbf{B}_2$.

\subsection{Optimal location of flow measurements}
Our first goal is to optimize the location of the edges to be sampled in order to find a suitable tradeoff between number of samples and reconstruction error. we leverage the Max-Det method introduced in \cite{tsitsvero2016signals} for node signals and extended to edge signals defined on a cell complex in \cite{sardellitti2024topological}. To find this set, we use the edge flows $\bx^{1}$ generated by the simulator EPANET. Let us denote by $\mathcal{F}$ the set of edges where the sensors should be placed to collect water flow measurements. To run the Max-Det algorithm, we first need to find a band-limited representation, possibly approximate, of the edge flow. We denote by $\bar{\mathbf{f}}_{\mathcal{F}}$ the vector of dimension $E$ whose entries $f_{\mathcal{F}}(i)$ are the flow measurements, for $i \in \mathcal{F}$, and zero elsewhere. To find the bandlimited representation, we can either use the eigenvectors of $\mL_{1, d}$, as in GSP-based approach working with the edge Laplacian, or with the eigenvectors of $\mL_1$, which is associated to the second order cell complex. In either case, we wish to find the set $\mathcal{M}$ of indexes of the eigenvectors whose linear combination best approximates the edge flow vector. More specifically, denoting with $\mU_1$ the eigenvectors of either $\mL_{1, d}$ or of $\mL_1$, we solve problem $\mathcal{P}_B$ in (\ref{eq:bas_pur}), where we set $\mV:= \mU_1$ and $\bx^{0}=\bx^{1}$. We denote by $\mU_{1,\mathcal{M}}$ the set of eigenvectors that provide a suitable trade-off between reconstruction error and sparsity of the representation. 
Using $\mU_{1,\mathcal{M}}$, we run the Max-Det algorithm and find the optimal sampling set $\cal{F}$.

\begin{figure}[ht]
    \centering
    {\includegraphics[width=1\columnwidth]{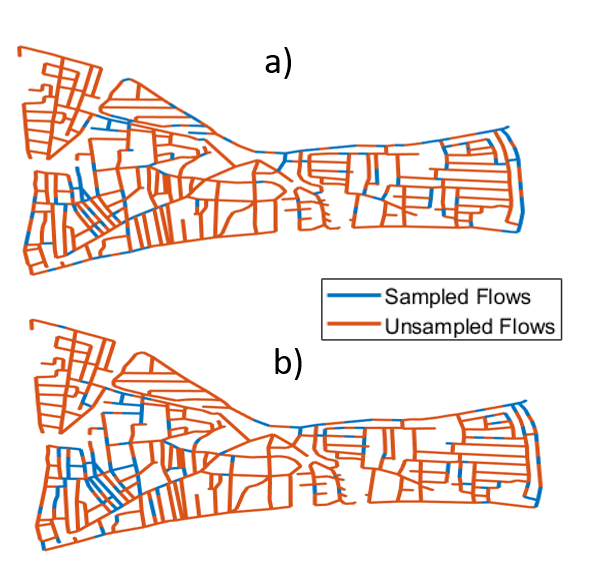}}
    \caption{Graph representation of  the L-town WDN, where we  colored in blue the sampled edges,  corresponding to edges equipped by measuring devices, while in orange the remaining nodes. In panel a) we have the scenario correspondent to the use of  $\mL_1$ bases while in panel b) we illustrate sampling obtained by using  the eigenvectors of  $\mL_{1,d}$,}
    \label{fig:flowsampling}
\end{figure}

To numerically test the effectiveness of the proposed strategy, we consider the L-town network represented in Fig. \ref{fig:flowsampling}.
The optimal sets $\mathcal{F}$ of edge samples obtained in the two cases are drawn in blue in Fig. \ref{fig:flowsampling}, while all other edges are in blue. 
The upper plot a) reports the sampling configuration corresponding to the case where the sampling algorithm is run using the eigenvectors of $\mL_1$ as a basis. The lower plot b) of Fig. \ref{fig:flowsampling}, reports the sampling strategy obtained using as basis the eigenvectors $\mU_d$ of the lower Laplacian matrix $\mL_{1,d}$, which represents the GSP-based approach.
Comparing plots a) and b), we can notice a slight change of the sampling strategy. This is not entirely surprising, because many eigenvectors of $\mL_{1}$ coincide with those of $\mL_{1,d}$. Nevertheless, this change has a non-negligible impact on the reconstruction of the overall flow vector from its samples, as shown in the next section.

\section{Reconstruction of the overall flow vector from its samples}
Given the sampling set $\cal{F}$, we build the diagonal selection matrix $\mD_{\cal{F}}$, whose diagonal $i$-th entry is either $1$, if $i \in \cal{F}$ or $0$ otherwise, and its complement $\bar{\mD}_{\mathcal{F}}:=\mathbf{I}-\mD_{\cal{F}}$. Then, we use the basis $\mU_{1,\mathcal{M}}$ found in the previous section to recover the flow over the unobserved edges, by solving the following optimization problem:
\begin{equation}
\label{eq:optim_probflow}
\begin{array}{lll}
\underset{\mathbf{f} \in \mathbb{R}^{E}}{\min}  & \parallel \!(\mI-\bar{\mD}_{\mathcal{F}} \mU_{1,\mathcal{M}} \mU_{1,\mathcal{M}}^T)\mathbf{f}-\bar{\mathbf{f}}_{\mathcal{F}}  \parallel^2  +\hspace{1cm} (\mathcal{P}_f)\\
& + \beta \parallel \mB_1  \mathbf{f} -\mathbf{d}\parallel^2\medskip\\
\text{s.t.}  &  \; \;   \;\mathbf{f} \geq \mathbf{0}
\medskip\\
\end{array}
\end{equation}
where: i) the first term in the objective function captures the data-fitting error that occurs when an edge signal is reconstructed from a restricted subset of samples; and ii) the second term enforces compliance with the flow conservation principle in the WDN, controlled by the non-negative penalty coefficient  $\beta\geq 0$. The constraint ensures that edge flow values are non-negative. 

Problem $\mathcal{P}_f$ is a strongly convex problem and we can derive its optimal solution in closed form by solving the KKT conditions associated with $\mathcal{P}_f$. More specifically, introducing the Lagrangian multipliers vector $\boldsymbol{\eta}\in\mathbb{R}^{E}_{++}$, the Lagrangian function of $\mathcal{P}_{f}$ is expressed as
\begin{equation}
\begin{split}
&\mathcal{L}(\mf,\boldsymbol{\eta})=\!\parallel(\mI-\bar{\mD}_{\mathcal{F}} \mU_{1,\mathcal{M}} \mU_{1,\mathcal{M}}^T)\mathbf{f}-\bar{\mathbf{f}}_{\mathcal{F}}  \parallel^2 + \beta \parallel \mB_1  \mathbf{f} -\mathbf{d}\parallel^2\\
&-\boldsymbol{\eta}^T\mathbf{f}.
\end{split}
\end{equation}
Hence, there exist a multiplier $\eta^{\star}(k)$ such that the optimal flow vector ${\mf}^{\star}$ satisfies the following Karush-Kuhn-Tucker (KKT) conditions of the convex problem $\mathcal{P}_f$: 
\begin{equation}
\label{eq:KKTf}
\begin{split}
& \text{(i)} \; \nabla_{\mf}(\mf,\boldsymbol{\eta})= \mW^{T} \mW \mf^{\star}-\mW^{T} \bar{\mathbf{f}}_{\mathcal{F}} + \beta \mB_1^T \mB_1 \mathbf{f}^{\star} - \beta \mB_1^T \mathbf{d}\medskip\\ &  - \frac{\boldsymbol{\eta}}{2}=\mathbf{0}\medskip\\
& \text{(ii)} \; \eta^{\star}(k) \geq 0, \;  \eta^{\star}(k) \cdot {f}^{\star}(k) =0, \quad  k=1,\ldots,E \medskip\\
& \text{(iii)} \; \mf^{\star} \geq \mathbf{0}
\end{split}
\end{equation}
where $\mW=(\mI-\bar{\mD}_{\mathcal{F}} \mU_{1,\mathcal{M}} \mU_{1,\mathcal{M}}^T)$.
The optimal multiplier $\boldsymbol{\eta}^{\star}$ can be derived using the conditions (ii) and (iii). Specifically, observe from (ii) that, if ${f}^{\star}(k)>0$, then $\eta^{\star}(k)=0$, so that the complementary slackness conditions in {\text{(ii)} are satisfied.
On the other hand, if  ${f}^{\star}(k)\leq 0$, to satisfy the constraint in {\text{(iii)}, the optimal solution is given by ${f}^{\star}(k)=0$. Therefore, from the first equation {\text{(i)}, we easily get the optimal closed form solution
\begin{equation}
\label{eq:f_2}
\begin{split}
&\mathbf{f}^{\star}=\text{max}\left[( \mW^{T} \mW +  \beta \mB_1^T \mB_1 )^{-1}\left(\mW^{T} \bar{\mathbf{f}}_{\mathcal{F}}+\beta \mB_1^T \mathbf{d}\right),\mathbf{0}\right].
\end{split}
\end{equation}

\noindent \textbf{Numerical results}

    


In this section, we compare the reconstruction capabilities of a GSP-based method relying only on the graph representation of the WDN with our method using a cell complex-based representation. To compare our method with the GSP-based approach, we run problem $\mathcal{P}_B$ in (\ref{eq:bas_pur}), assuming as a basis matrix $\mV$ the eigenvector matrix $\mU_{d}$ of either the lower Laplacian matrix $\mL_{1,d}$ or of the Hodge Laplacian $\mL_1$, in order to find a sparse edge signal representation in the two cases. In the cell complex case, we build the incidence matrix $\mB_2$ including all polygonal cells with a maximum number of sides equal to $P_{max}=30$ 
Then, we solve the convex problem $\mathcal{P}_f$ in (\ref{eq:optim_probflow}) to recover the entire water flow vector on the network illustrated in Fig. \ref{fig:flowsampling} from the observed samples, setting the regularization parameter $\beta$ equal to $0.3$. Finally, we compute the MSE between the reconstructed flow signal and the ground truth signal, varying the number of known samples from $100$ to $600$. 
\begin{figure}[ht]
    \centering
    {\includegraphics[width=1\columnwidth]{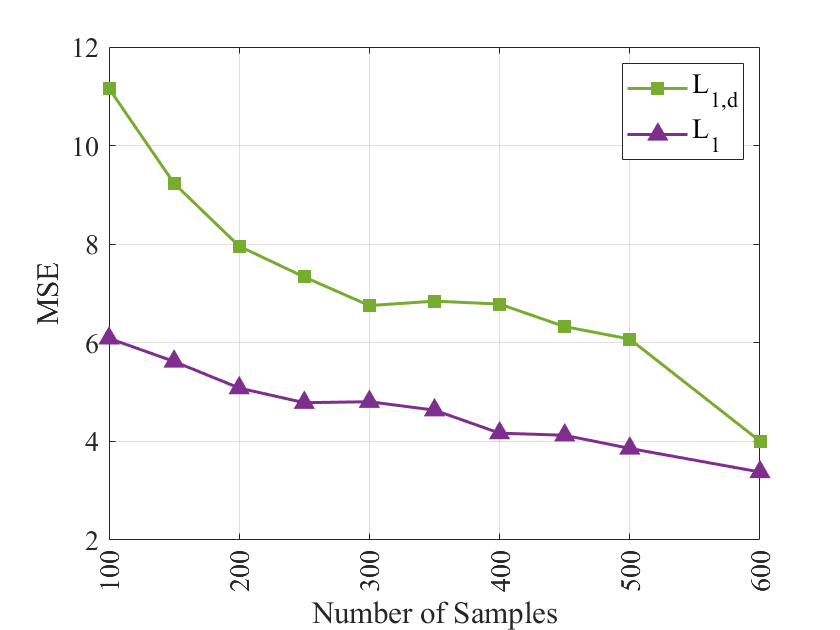}}
    \caption{MSE versus  the number of samples used to reconstruct the signal using cell-complexes eigenvector bases or graph eigenvectors  bases.} 
    \label{fig:mseL1}
\end{figure}

The result is reported in Fig. \ref{fig:mseL1}, where we show the MSE versus the number of samples, i.e. the number of non-zeros entries of the observed vector $\bar{\mathbf{f}}_{\mathcal{F}}$. We can notice how the cell complex-based approach yields a lower MSE, for any number of samples, with respect to the GSP-based approach. This result corroborates the idea that the cell complex topology is better able to capture the relations between flow signals, with respect to a simple graph structure. In particular, the inclusion of the term $\mL_{1, u}:=\mB_2 \mB_2^T$ gives rise to eigenvectors that properly encode the flow circulation along the boundary of each polygon (cell), so that including these vectors in the basis to be used to select the best representation of the overall flow vector provides a clear advantage. 
It is also interesting to note how the performance gain achieved using the cell complex-based representations is higher when the number of samples is low, while it decreases as the number of samples increases. This is highly relevant in practice, where the number of sensors is a critical limitation.

The results shown in Fig. \ref{fig:mseL1} refer to the network sketched in Fig. \ref{fig:flowsampling}. To test how the above results remain valid also under a very different network structure, we also used the WDN \cite{kydataset} sketched in Fig.\ref{fig:ky5} a), composed of $421$ junctions, $4$ reservoirs, $3$ tanks, $495$ pipes and $9$ pumps. 
\begin{figure}[ht]
    \centering
    \subfloat[]{\includegraphics[width=0.7\columnwidth]{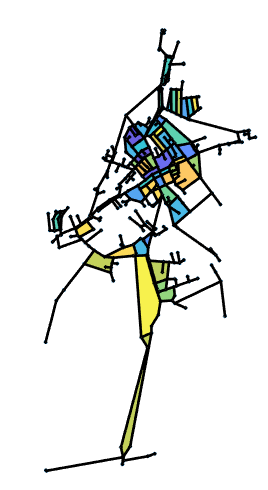}}\\ 
    \subfloat[]
    {\includegraphics[width=1\columnwidth]{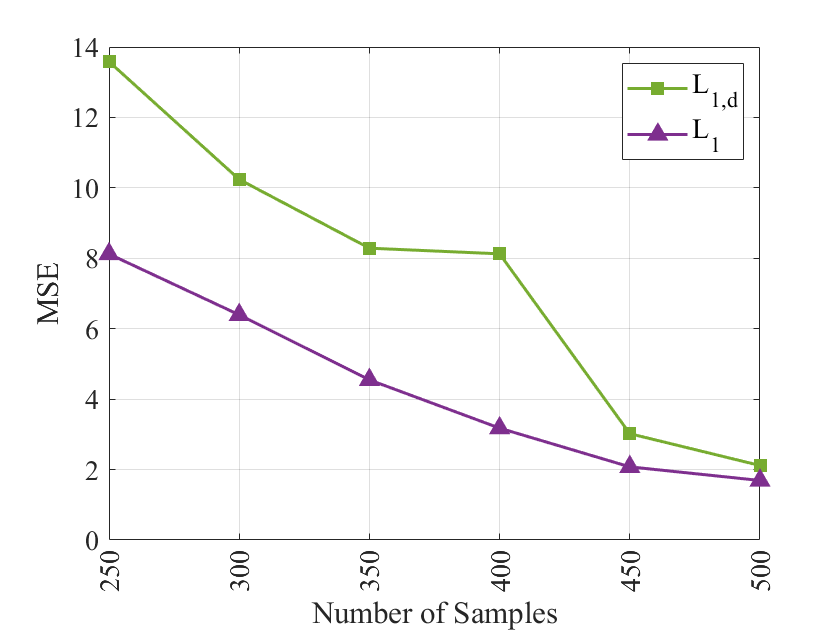}}
    \caption{In the upper plot (a) we represent  the WDN in \cite{kydataset} , where each polygonal cell is highlighted in color. In the bottom plot (b), we report the MSE versus the number of samples. The violet curve refers to the cell-based representation while the green curve to graph-based representation.} 
    \label{fig:ky5}
\end{figure}
This WDN was originally used by \cite{jolly2014research} for classification purposes. In Fig. \ref{fig:ky5} a) we also represent the cell complex associated with this WDN, incorporating $2$-order cells  for each polygon having a number of sides up to \( P_{\text{max}} = 10 \). The different colors represent the different cells. This network has been processed using EPANET to simulate water flow dynamics as before. Subsequently, we sample some of the edges using the same strategy used above and reconstruct the whole flow from its samples, using a GSP-based approach or our approach. In these simulations, the regularization parameter \( \lambda \) was set to $0.2$. The result is reported in Figure \ref{fig:ky5}b), which shows the MSE vs. the number of samples for cell- and graph-based learning strategies. We can check that, also in this case, our method provides better performance with respect to the GSP-based approach.\\  

\begin{figure}[ht]
    \centering
    {\includegraphics[width=1\columnwidth]{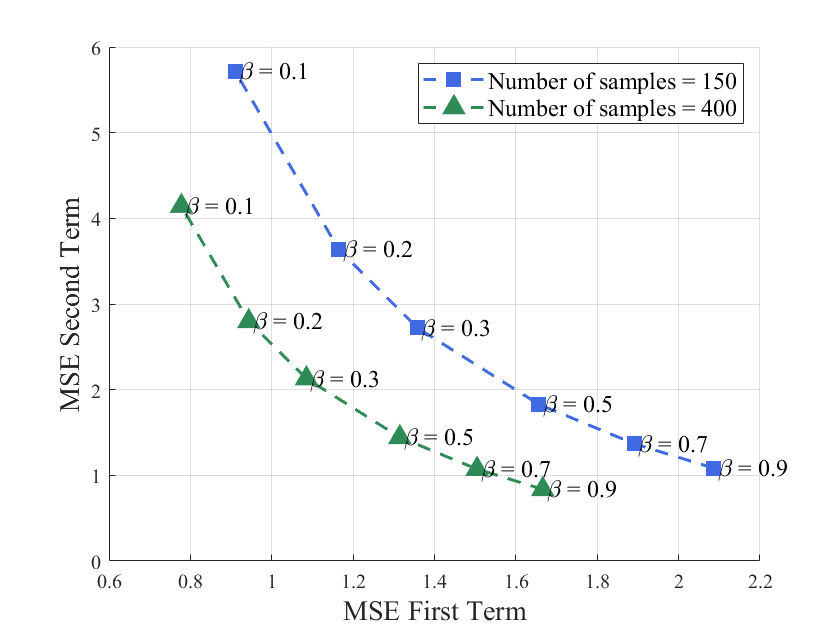}}
    \caption{Representation of the MSE of the two terms of the minimization problem $\mathcal{P}_f$ as the parameter $\beta$ varies. The first MSE term (x-axis) and the second MSE term (y-axis) are shown for two conditions: with $150$ samples (depicted by a dashed line with square markers in light blue) and $400$ samples (depicted by a dashed line with triangle markers in green). Each point corresponds to a specific value of the regularization parameter $\beta$, annotated near to the respective point.}
    \label{fig:msefirstsecond}
\end{figure}

Finally, to explore the trade-off  between the two terms appearing in the objective function of the optimization problem $\mathcal{P}_f$, in Fig.   \ref{fig:msefirstsecond} we report the MSE corresponding to the violation of the flow conservation law at the nodes (second term in \eqref{eq:optim_probflow}) versus the data fitting error (first term in \eqref{eq:optim_probflow}, for different values of the parameter $\beta$. Specifically, we represent the simulation results using $150$ flow samples (dashed line with square markers in light blue) and $400$ samples (dashed line with triangle markers in green). Each point corresponds to a specific value of the regularization parameter $\beta$, as annotated near the respective point. 
From Fig. \ref{fig:msefirstsecond}, we can observe how, increasing the regularization parameter $\beta$, we give more weight to the flow conservation law with respect to the flow reconstruction error. Furthermore, as expected, we can observ the performance gain obtained by increasing the number of sampled flows.\\


Therefore, the results obtained from the previous datasets demonstrate that reconstructing water flow values from a limited subset of measurements is more effective when leveraging the higher-order Laplacian within a topological signal processing framework. This approach ensures that higher-order interactions, derived from the presence of cell complexes in closed graph structures, are not neglected, and they enhance edge signal reconstruction accuracy by accounting for flows circulations along the polygonal cells. These circulations are instrumental in identifying cycles where energy conservation  laws are violated.

\section{Conclusion}
\label{sec:concl}
In this paper we have introduced a cell complex-based approach to better represent the relationships between node and edge variables in a water distribution network. In particular, we merged a physics-based approach, incorporating mass and energy conservation laws, with topological signal processing methodologies, to extract a better representation of the pressure and flow variables across the network. More specifically, we incorporated a nonlinear relation between pressure and flow, to take into account friction effects arising when water flows through a real pipe. Building on this physics-informed topological signal processing approach, we proposed a novel way to optimally monitor a WDN and reconstruct the whole state of pressures and flows from a limited number of sensor measurements. 
The proposed recovery methods have been validated on realistic WDNs, obtained from real networks, and using the EPANET software to generate realistic pressures and flows across the network. We evaluated the potential of the framework to handle practical challenges such as limited sensor deployments, nonlinear flow-pressure relationships, and the influence of active elements. 

Our results demonstrate that the inclusion of cell complexes and higher-order Laplacians plays a key role in capturing the complex interactions between node and flow variables, in particular to represent water flow circulations along the cycles of the network. An interesting direction for future research is to generalize the proposed method to include the dynamics of the whole process and to possibly detect anomalies due, for example, to water leakages.

\ifCLASSOPTIONcaptionsoff
  \newpage
\fi

\bibliography{biblio}
\bibliographystyle{ieeetr}

\end{document}